\documentclass[12pt,aps,prd,showpacs,superscriptaddress,here]{revtex4}
\usepackage{graphicx}
\usepackage{bm}
\begin{document}
\title{Lagrangian perturbation theory in Newtonian cosmology}

\author{Takayuki Tatekawa}
\email{tatekawa@gravity.phys.waseda.ac.jp}
\affiliation{Department of Physics, Waseda University,
3-4-1 Okubo, Shinjuku-ku, Tokyo 169-8555, Japan}
%
\date{January 23, 2005}
\begin{abstract}
We discuss various analytical approximation methods for the evolution
of the density fluctuation in the Universe. From primordial density
fluctuation, the large-scale structure is formed via its own
self-gravitational instability. For this dynamical evolution,
several approaches have been considered. In Newtonian cosmology,
in which matter motion is described by Newtonian dynamics with cosmic
expansion, Lagrangian description for the cosmic fluid is known
to work rather well for quasi-nonlinear clustering regime.
In this paper, we
briefly review Lagrangian perturbation theory in Newtonian cosmology.
\end{abstract}

\pacs{04.25.Nx, 95.30.Lz, 98.65.Dx}

\maketitle

\section{Introduction}
There are various structures in the Universe. For example,
galaxies, group of galaxies, cluster of galaxies, 
voids, large-scale structure, and so on. How are such structures
formed? As the most plausible scenario, we have considered that
a primordial density fluctuation spontaneously produces by
itself a gravitational instability. The fluctuation could
have been
generated by several processes in the early Universe
\cite{Kolb90, Coles95, Liddle00, Linde90}. 
When we do not consider the superhorizon scale or extremely
dense region like a supermassive black hole,
the motion of the cosmological fluid can be described
by Newtonian cosmology. In Newtonian cosmology,
the motion of matter is described by Newtonian dynamics,
and the cosmic expansion is given by the Friedmann equation.

The Lagrangian description for the cosmological fluid
can be usefully applied to
the structure formation scenario. This description provides a
relatively accurate model even in a quasi-linear regime.
Zel'dovich~\cite{Zeldovich70} proposed
a linear Lagrangian approximation for dust fluid.
This approximation is called the Zel'dovich approximation
(ZA)~\cite{Zeldovich70,Arnold82,Shandarin89,Buchert89,Coles95,Sahni95}.
After that, higher-order approximation for the Lagrangian description
was proposed
\cite{Bouchet92,Buchert93,Buchert94,Bouchet95,Catelan95,Buchert92,Barrow93,Sasaki98}.
Then it was shown that the Lagrangian approximation
describes the evolution of density fluctuation better than
the Eulerian approximation~\cite{Munshi94,Sahni96,Yoshisato98}.
Especially for the planar model, ZA gives an exact solution
\cite{Doroshkevich73,Shandarin89}.

Although the Lagrangian approximation gives an accurate
description until a quasi-linear regime develops,
it cannot describe the model after the formation of caustics.
For example, in ZA, even after the formation of caustics,
the fluid elements
keep moving in the direction set up by the initial condition.
Therefore, the nonlinear structure that forms diffuses.

In order to proceed with a hydrodynamical description
in which caustics do not form, a qualitative pressure gradient
\cite{Zeldovich82}
and thermal velocity scatter~\cite{Kotok87, Shandarin89}
in a collisionless matter have been discussed. Additionally
the ``adhesion approximation'' (AA)~\cite{Gurbatov89} was proposed
based on the nonlinear diffusion equation
(Burgers' equation). In AA, an artificial
viscosity term is added to ZA. Because of the viscosity term,
we can avoid caustics formation. Using AA,
the problem of structure formation has been
discussed~\cite{Shandarin89,Weinberg90,Nusser90,Kofman92,Melott94A}.
The density divergence does not occur in AA, and the density distribution
close to the N-body simulation can be produced.
However, the origin of the viscosity has not yet been clarified.
Another modified model for the Lagrangian approximation is known
as the ``Truncated (or Optimized) Zel'dovich approximation'' (TZA)
\cite{Coles93, Melott94B}.
The evolution equation in TZA is the same as that in ZA. Instead of
the evolution, the initial power spectrum is changed in TZA.
Because the caustics are formed by small-scale fluctuation,
TZA introduces some cutoff in the small scale in the initial spectrum.
As a result, TZA avoids the formation of the caustic.
However, the physical meaning of the cutoff also
has not yet been clarified.

For another purpose, improvement of approximation, several
modified models have been proposed. For example, Bagla and
Padmanabhan~\cite{Bagla94} developed a new scheme that
extended the frozen
flow approximation~\cite{Matarrese92}. Then they showed better
agreement with the N-body simulations than ZA.
Instead of ordinary perturbative
expansion, the Pad\'{e} approximation was introduced
to Lagrangian approximation~\cite{Yoshisato98,Matsubara98}.
When we consider the spherical void evolution,
if we increase the order of
Lagrangian approximation to try to improve,
contrary to expectation, the description becomes
worse~\cite{Munshi94,Sahni96}.
On the other hand, the Pad\'{e} approximation improves
the description of the void evolution.
Another approach is generalized ZA. Although ZA solutions
are independent of position, the solutions of 
generalized models depend on position. 
Although the computation becomes complicated,
we can obtain exact solutions
for planar, cylindrical, and spherical symmetric models
\cite{Reisenegger95,Audit96,Betancort-Rijo00,Makler01}.

In ZA and its extended models, pressure was ignored.
Recently, Lagrangian approximation in which the effect of pressure
was taken into consideration have been analyzed.
Buchert and Dom\'{\i}nguez~\cite{Buchert98} discussed the effect
of velocity dispersion using the collisionless Boltzmann equation~\cite{Binney87}.
They argued that models of a large-scale structure should
be constructed for a flow describing the average
motion of a multi-stream system.
Then they showed that when the velocity dispersion is regarded
as small and isotropic it produces effective ``pressure'' or
viscosity terms. Furthermore, they derived the
relation between mass density $\rho$ and pressure $P$, i.e.,
an ``equation of state.''
Buchert \textit{et al.}~\cite{Buchert99} showed how the viscosity term
or the effective pressure of a fluid is generated,
assuming that the peculiar acceleration is
parallel to the peculiar velocity.
Dom\'{\i}nguez~\cite{Dominguez00,Dominguez02} clarified that a hydrodynamic
formulation is obtained via a spatial coarse-graining
in a many-body gravitating system, and that the viscosity term in
AA can be derived by the expansion
of coarse-grained equations.

With respect to the relation between the viscosity term
and effective pressure, and the extension of the Lagrangian description
to various matter,
the Lagrangian perturbation theory of pressure has been considered.
Adler and Buchert~\cite{Adler99} have formulated the
Lagrangian perturbation theory for a barotropic fluid.
Morita and Tatekawa~\cite{Morita01} and Tatekawa \textit{et al.}
\cite{Tatekawa02}
solved the Lagrangian perturbative equations for a polytropic fluid
up to the second order.
Hereafter, we call this model the ``pressure model.''
The behavior of the pressure model strongly depends on the polytropic
exponent $\gamma$~\cite{Morita01, Tatekawa02, Tatekawa04A}.
According to a recent study~\cite{Tatekawa04B}, the effect of pressure
cannot realize the artificial viscosity in AA completely.

Matarrese and Mohayaee proposed Lagrangian perturbative equation
for two component (baryon and cold dark matter) fluids
\cite{Matarrese02}. They consider lowest-order Lagrangian perturbation
for the two components and discussed the evolution of
the density fluctuation.

We especially mention only various Lagrangian perturbation models.
For advanced and related topics, we recommend to refer other
reviews and books
(\cite{Paddy93, Coles95, Sahni95, Peacock99, Jones04}, for example).

This paper is organized as follows.
In Sec.~\ref{sec:basic}, we present basic equations for the cosmological
fluid in Newtonian cosmology. First, we show Eulerian description for
the basic equations in Sec.~\ref{sec:Eulerian}. Then we introduce
Lagrangian description in Sec.~\ref{sec:Lagrangian}.
In Sec.~\ref{sec:dust}, we show perturbative solutions for dust fluid
up to a third-order approximation. First, we consider first-order longitudinal
solutions in Sec.~\ref{subsec:ZA}. The first-order approximation is called
``Zel'dovich approximation'' (ZA). After that, a higher-order
approximation is proposed. In Sec.~\ref{subsec:PZA},
we mention second- and third-order longitudinal
solutions. In Sec.~\ref{subsec:Trans}, we explain the transverse mode.
When we analyze
vorticity for cosmological fluid, we need to consider this mode.
Although these solutions seem useful, they do not have physical
meaning after formation of a caustic. To avoid formation of a caustic,
several
modified models have been proposed. In Sec.~\ref{subsec:modify1} and
\ref{subsec:modify2}, we show
modified models for Lagrangian perturbation. In Sec.~\ref{subsec:modify1},
we explain some modification to avoid the formation of caustics.
In Sec.~\ref{subsec:modify2}, we show alternative models to obtain
high accuracy.
In Sec.~\ref{sec:pressure},
we explain the pressure model.
In Sec.~\ref{sec:summary}, we summarize the success and failure
in Lagrangian approximation and state applications.
We notice the spherical void evolution with dust fluid
in Appendix~\ref{sec:void}. {\em In Appendix~\ref{sec:quantities},
we provide a the table explaining the physical quantities used
in this paper.}

\section{Basic equations}\label{sec:basic}
\subsection{Basic equations for the cosmological fluid}\label{sec:Eulerian}

In this section, we present basic equations for the cosmological fluid
in Newtonian cosmology; i.e., the motion of fluid is described by
Newtonian dynamics, and the background cosmic expansion is given
by the Friedmann equation. In this situation, the motion of
perfect fluid
is described by a continuous equation, Euler's equation,
and Poisson's equation~\cite{Peebles81,Coles95,Sahni95}.
\begin{eqnarray}
\left (\frac{\partial \rho}{\partial t} \right )_r + \nabla_r \cdot
(\rho \bm{u}) &=& 0 \,, \label{eqn:continuous} \\
\left (\frac{\partial \bm{u}}{\partial t} \right )_r + (\bm{u} \cdot
\nabla)_r \bm{u} &=& - \frac{1}{\rho} \nabla_r P + \bm{g} \,,
\label{eqn:Euler-eq} \\
\bm{g} = - \nabla_r \Phi\,, \nabla \cdot \bm{g} &=& 4 \pi G \rho
\,, \label{eqn:Poisson}
\end{eqnarray}
To introduce cosmic expansion, we adopt the coordinate transformation
from physical coordinates to comoving coordinates.
\begin{equation} \label{eqn:comoving-coordinate}
\bm{x} = \frac{\bm{r}}{a(t)}, a(t):\mbox{scale factor} \,,
\end{equation}
where $\bm{r}$ and $\bm{x}$ are physical coordinates and
comoving coordinates, respectively.
The scale factor satisfies the Friedmann equations:
\begin{eqnarray}
\left ( \frac{\dot{a}}{a} \right )^2 &=&
H^2 = \frac{8 \pi G}{3} \rho_b - \frac{\mathcal{K}}{a^2}
 + \frac{\Lambda}{3} \,, \label{eqn:Friedmann-00} \\
\frac{\ddot{a}}{a} &=& -\frac{4 \pi G}{3} \rho_b
 + \frac{\Lambda}{3} \,, \label{eqn:Friedmann-ii} \\
H & \equiv & \frac{\dot{a}}{a} \label{eqn:def-Hubble} \,,
\end{eqnarray}
with a curvature constant $\mathcal{K}$ and a cosmological
constant $\Lambda$. 
$H = \dot{a}/a$ and $\rho_b$ are Hubble parameter
and background density, respectively. 
Here we define density parameter $\Omega$:
\begin{equation}
\Omega \equiv \frac{8 \pi G}{3 H^2} \rho_b \,.
\end{equation}
Under the transformation (\ref{eqn:comoving-coordinate}),
the velocity and partial differential operator are changed as
\begin{eqnarray} \label{eqn:differential-ope}
\bm{u} = \dot{\bm{r}} = \dot{a} \bm{x} + \bm{v}(\bm{x}, t) ,~~~ (\bm{v}
\equiv a \dot{\bm{x}}) \,, \\
\nabla_x = a \nabla_r  \,, \\
\left (\frac{\partial f(\bm{x}=\bm{r}/a, t)}{\partial t} \right )_r =
\left (\frac{\partial f}{\partial t} \right )_x - \frac{\dot{a}}{a}
(\bm{x} \cdot \nabla_x) f \,.
\end{eqnarray}
Here we define the density fluctuation $\delta$ as follows:
\begin{equation} \label{eqn:def-delta}
\rho=\rho_b(t) \{ 1+ \delta(\bm{x}, t) \} \,, 
~~~ \rho_b \propto a^{-3} \,.
\end{equation}
In the comoving coordinates,
the basic equations for cosmological fluid are described as
\begin{eqnarray}
\frac{\partial \delta}{\partial t} + \frac{1}{a} \nabla_x \cdot \{ \bm{v}
(1+\delta) \} &=& 0 \,, \label{eqn:comoving-conti-eq} \\
\frac{\partial \bm{v}}{\partial t} + \frac{1}{a} (\bm{v} \cdot \nabla_x)
\bm{v} + \frac{\dot{a}}{a} \bm{v} &=& \frac{1}{a} \tilde{\bm{g}} - \frac
{1}{a \rho} \nabla_x P \,, \label{eqn:comoving-Euler-eq} \\
\nabla_x \times \tilde{\bm{g}} &=& \bm{0} \,, \label{eqn:rot-g} \\
\nabla_x \cdot \tilde{\bm{g}} &=& - 4 \pi G \rho_b a \delta \,.
\label{eqn:comoving-Poisson-eq}
\end{eqnarray}
Therefore, we obtain from Eqs.~(\ref{eqn:comoving-Euler-eq}) and
(\ref{eqn:rot-g}),
\begin{equation} \label{eqn:rot-Euler-eq}
\nabla_x \times \left (\frac{\partial \bm{v}}{\partial t}
 + \frac{\dot{a}}{a} \bm{v} + \frac{1}{a} \left ( \bm{v} \cdot \nabla_x 
 \right ) \bm{v} \right ) = \bm{0} \,,
\end{equation}
and from Eqs.~(\ref{eqn:comoving-Euler-eq}) and
(\ref{eqn:comoving-Poisson-eq}),
\begin{equation} \label{eqn:div-Euler-eq}
\nabla_x \cdot \left (\frac{\partial \bm{v}}{\partial t}
 + \frac{\dot{a}}{a} \bm{v} + \frac{1}{a} \left ( \bm{v} \cdot \nabla_x 
 \right ) \bm{v} \right ) =-4\pi G \rho_b \delta - \nabla_x \cdot
\left ( \frac{1}{\rho a}  \nabla_x P \right ) \,.
\end{equation}
%
\subsection{The Lagrangian description for the cosmological fluid}
\label{sec:Lagrangian}

In the Lagrangian hydrodynamics,
the coordinates $\bm{x}$ of the fluid elements are
represented in terms of Lagrangian coordinates $\bm{q}$ as
\begin{equation} \label{eqn:Euler-Lagrange}
\bm{x} = \bm{q} + \bm{s} (\bm{q},t) \,,
\end{equation}
where $\bm{q}$ is defined as initial values of $\bm{x}$,
and $\bm{s}$ denotes the Lagrangian displacement vector
due to the presence of inhomogeneities. The peculiar velocity is given by
\begin{equation}
\bm{v}=a \dot{\bm{s}} \label{eqn:L-velocity} \,.
\end{equation}
Then we introduce the Lagrangian time derivative:
\begin{equation}
\frac{\rm d}{{\rm d} t} \equiv \frac{\partial}{\partial t}
 + \frac{1}{a} \bm{v} \cdot \nabla_x \,.
\end{equation}
Using the Lagrangian derivative, we rewrite Eqs.~(\ref{eqn:rot-Euler-eq})
and (\ref{eqn:div-Euler-eq}).
The nonlinear term of the peculiar velocity disappears.
\begin{eqnarray} 
\nabla_x \times \left (\frac{{\rm d} \bm{v}}{{\rm d} t}
 + \frac{\dot{a}}{a} \bm{v} \right )
 &=& \bm{0} \label{eqn:rot-Euler-eqL} \,, \\
\nabla_x \cdot \left (\frac{{\rm d} \bm{v}}{{\rm d} t}
 + \frac{\dot{a}}{a} \bm{v} 
 \right ) &=& -4\pi G \rho_b \delta - \nabla_x \cdot
\left ( \frac{1}{\rho}  \nabla_x P \right ) \,.
\label{eqn:div-Euler-eqL}
\end{eqnarray}

The exact form of the energy density in the Lagrangian space
is obtained from Eq.~(\ref{eqn:comoving-conti-eq}) as
\begin{equation}\label{eqn:exactrho}
\rho = \rho_{\rm b} J^{-1} \,,
\end{equation}
where $J \equiv \det (\partial x_i / \partial q_j)
= \det (\delta_{ij} + \partial s_i / \partial q_j)$
is the Jacobian of the coordinate transformation from
$\bm{x}$ to $\bm{q}$. $J$ is described by expansion of
the derivative of Lagrangian perturbation as follows:
\begin{equation} \label{eqn:L-Jacobian}
J = 1+ \nabla_q \cdot \bm{s} +
 \frac{1}{2} \left ( (\nabla_q \cdot \bm{s})^2 
 - \frac{\partial s_i}{\partial q_j}
  \frac{\partial s_j}{\partial q_i} \right )
 + \mbox{det} \left ( \frac{\partial s_i}{\partial q_j}
 \right ) \,.
\end{equation}
Next, we consider the description of the
vorticity in Lagrangian space. We define the vorticity as
\begin{equation}
\bm{\omega} \equiv \frac{1}{a} \nabla_x \times \bm{v} \,.
\end{equation}
From Eq.~(\ref{eqn:rot-Euler-eqL}), we obtain the vorticity equation.
\begin{equation} \label{eqn:vorticity}
\frac{{\rm d} \bm{\omega}}{{\rm d} t} + 2 \frac{\dot{a}}{a} \bm{\omega}
 + \frac{\bm{\omega}}{a} \left ( \nabla_x \cdot \bm{v} \right )
= \left ( \bm{\omega} \cdot \nabla_x \dot{\bm{x}} \right ) \,.
\end{equation}
In the Lagrangian description, Eq.~(\ref{eqn:vorticity}) is solved
exactly.
\begin{equation}
\bm{\omega} = \frac{\left(\bm{\omega}_0 (\bm{q}) \cdot \nabla_q \right )
 \bm{x} ({\bm q}, t)}{a^2 J} \,,
\end{equation}
where $\bm{\omega}_0 ({\bm q})$ is the primordial vorticity.

From Eqs.~(\ref{eqn:L-velocity}) and (\ref{eqn:div-Euler-eqL}),
the peculiar gravitational field is written as
\begin{equation} \label{eqn:Lagrangian-Euler}
\tilde{\bm{g}} = a \left(
\ddot{\bm{s}}
+ 2\frac{\dot a}{a} \dot{\bm{s}}
- \frac{1}{a^2} \frac{{\rm d} P}{{\rm d} \rho} (\rho)
  \, J^{-1} \nabla_{\bm{x}} J \right) \,,
\end{equation}
where an overdot $(\dot{\mbox{ }})$ denotes ${\rm d}/{\rm d} t$.
Hence, from Eqs.~(\ref{eqn:rot-g}) and (\ref{eqn:comoving-Poisson-eq}),
we obtain the following equations for $\bm{s}$:
\begin{equation}\label{rot-ddots}
\nabla_{\bm{x}} \times
\left( \ddot{\bm{s}}
+ 2\frac{\dot a}{a} \dot{\bm{s}}
\right) = 0 \,,
\end{equation}
\begin{equation}\label{div-ddots}
\nabla_{\bm{x}} \cdot \left(
\ddot{\bm{s}}
+ 2\frac{\dot a}{a} \dot{\bm{s}}
- \frac{1}{a^2} \frac{{\rm d} P}{{\rm d} \rho}
  \, J^{-1} \nabla_{\bm{x}} J \right)
= -4\pi G\rho_{\rm b} (J^{-1} -1) \,.
\end{equation}
If we find solutions of Eqs.~(\ref{rot-ddots})
and (\ref{div-ddots}) for $\bm{s}$,
the dynamics of the system considered is completely determined.
Since these equations are highly nonlinear and hard to solve exactly,
we will advance a perturbative approach.
Remark that, in solving the equations for $\bm{s}$
in the Lagrangian coordinates $\bm{q}$,
the operator $\nabla_{\bm{x}}$ will be transformed
into $\nabla_{\bm{q}}$
by the following rule:
\begin{equation}
\frac{\partial}{\partial q_i} =
\frac{\partial x_j}{\partial q_i} \frac{\partial}{\partial x_j}
= \frac{\partial}{\partial x_i} +
\frac{\partial s_j}{\partial q_i} \frac{\partial}{\partial x_j} 
\,.
\end{equation}

We decompose $\bm{s}$ into the longitudinal
and the transverse modes as
$\bm{s} = \nabla_{\bm{q}} S
+ \bm{S}^{\rm T}$ with
$\nabla_{\bm{q}} \cdot \bm{S}^{\rm T}=0$.

%
\section{Dust model}\label{sec:dust}
\subsection{Zel'dovich approximation}\label{subsec:ZA}

Zel'dovich derived a first-order solution of the longitudinal mode
for dust fluid~\cite{Zeldovich70}. The evolution equation for
first-order solution is written as
\begin{equation}
\ddot{S}^{(1)} + 2 \frac{\dot{a}}{a} \dot{S}^{(1)}
 - 4\pi G \rho_b S^{(1)} =0 \,.
\end{equation}
Then, the first-order solutions are written as follows:
\begin{equation} \label{eqn:sol-ZA}
S^{(1)} (\bm{q}, t) = D_+(t) \psi^{(1)}_+ (\bm{q})
 + D_-(t) \psi^{(1)}_- (\bm{q}) \,.
\end{equation}
For the E-dS Universe model ($\Omega =1$), $D_{\pm}$ is written as
\begin{eqnarray}
D_+(t) &=& t^{2/3} \propto a(t) \,, \\
D_-(t) &=& t^{-1} \propto a(t)^{-3/2} \,.
\end{eqnarray}
For a non-flat model, following Doroshkevich \textit{et al.}
\cite{Doroshkevich73}, we replace the standard cosmological
time by a new time $\tau$ defined by
\begin{equation} \label{eqn:def-newtime}
{\rm d} \tau \propto a^{-2} {\rm d} t \,.
\end{equation}
Using new time, we obtain an analytic form for $D_{\pm}$.
For the open model ($\Omega <1$),
\begin{eqnarray} \label{eqn:LA-open-1st}
D_+ (\tau) &=& 1+ 3(\tau^2-1) \left (1+ \frac{\tau}{2}
 \log \left ( \frac{\tau-1}{\tau +1} \right ) \right ) \,,
\\
D_- (\tau) &=& \tau (\tau^2 -1) \,.
\end{eqnarray}
For the closed model ($\Omega >1$), we obtain the analytic form
by the transformation
$\tau \rightarrow i \tau$ in Eq.~(\ref{eqn:LA-open-1st}).

For $\Lambda$-flat model, in which the cosmological constant exists
and the Universe is flat, $D_{\pm}$ is written as
\begin{eqnarray}
D_+ (t) &=& h \int_h^{\infty} \frac{{\rm d} \theta}{\theta^2
 (\theta^2-1)} = \frac{h}{2} B_{1/h^2} \left(\frac{5}{6},
 \frac{2}{3} \right ) \,, \\
D_- (t)&=& h \,, \\
h &=& \frac{H(t)}{\sqrt{\Lambda/3}} \,,
\end{eqnarray}
where $B_{1/h^2}$ is incomplete Beta function:
\begin{equation}
B_z (\mu, \nu) \equiv \int_0^z p^{\mu-1} (1-p)^{\nu-1} {\rm d} p \,.
\end{equation}
This first-order approximation is called the Zel'dovich
approximation (ZA). Especially when we consider
the plane-symmetric case, ZA gives exact solutions
\cite{Doroshkevich73,Arnold82}.

Using Eqs.~(\ref{eqn:exactrho}) and (\ref{eqn:sol-ZA}),
we can describe density fluctuation. Here we define
the deformation tensor $X_{\alpha \beta}$:
\begin{equation} \label{eqn:ZA-deformation}
X_{\alpha \beta} \equiv \frac{\partial x_{\alpha}}{\partial q_{\beta}}
= \delta_{\alpha \beta} + \frac{\partial s_{\alpha}}{\partial q_{\beta}}
\,.
\end{equation}
The eigenvalues of the deformation tensor (\ref{eqn:ZA-deformation})
are written by
\begin{equation} \label{eqn:ZA-eigenvalue}
w_i = -D(t) \lambda_i^0 (\bm{q}) \,,
\end{equation}
where $\lambda_i^0$ is the eigenvalue of $\partial \psi_{,\alpha}/\partial
q_{\beta}$.
Using these eigenvalues, the density fluctuation is described
by
\begin{equation} \label{eqn:delta-eigenvalue}
(1+\delta)^{-1} = \Pi_{i=1}^3 (1+w_i) \,.
\end{equation}
When $w_i$ becomes $-1$,
the caustic (``Zel'dovich's pancake'') is formed and the
density fluctuation diverges. After that, the perturbative solution
does not have physical meaning. In order to avoid the formation
of caustics, several modified models have been proposed. We will 
discuss these models in Sec.~\ref{subsec:modify1}. By generalizing
the eigenvalues of the deformation tensor
(Eq.~(\ref{eqn:ZA-eigenvalue})),
more exact approximations can be proposed. These models will
be expressed in Sec.~\ref{subsec:modify2}.

\subsection{Higher-order approximation}\label{subsec:PZA}

ZA solutions are known as perturbative solutions, which describe
the structure
well in the quasi-nonlinear regime. To improve approximation,
higher-order perturbative solutions of Lagrangian displacement
were derived.
Irrotational second-order solutions (PZA) were derived by
Bouchet \textit{et al.}~\cite{Bouchet92} and Buchert and Ehlers~\cite{Buchert93},
and third-order solutions (PPZA) were obtained by Buchert~\cite{Buchert94},
Bouchet \textit{et al.}~\cite{Bouchet95}, and Catelan~\cite{Catelan95}.

Hereafter we consider only growing-mode $D_+$
in first-order Lagrangian perturbation.
First, we show the evolution equation for second- and third-order
perturbative equations for the longitudinal mode. Here we decompose the
Lagrangian perturbation to time and a spacial component.
\begin{eqnarray}
S^{(2)} (t, \bm{q}) &=& E(t) \psi^{(2)} (\bm{q}) \,, \\
S^{(3)} (t, \bm{q}) &=& F(t) \psi^{(3)} (\bm{q}) \,,
\end{eqnarray}
where the superscript $S^{(n)}$ means n-th order Lagrangian perturbation.
The second-order perturbative equation is written as
\begin{eqnarray}
&& \left (\ddot{E} + 2 \frac{\dot{a}}{a} \dot{E}
- 4 \pi G \rho_b E \right ) \psi_{,ii}^{(2)} \nonumber \\
&=& -2 \pi G \rho_b D^2 \left[ (\psi_{,ii}^{(1)})^2
 - \psi_{,ij}^{(1)} \psi_{,ji}^{(1)} \right ] \,.
 \label{eqn:Lagrange-2ndL}
\end{eqnarray}
The third-order perturbative equation becomes little
complicated.
\begin{eqnarray}
&& \left (\ddot{F} + 2 \frac{\dot{a}}{a} \dot{F}
- 4 \pi G \rho_b F \right ) \psi_{,ii}^{(3)} \nonumber \\
&=& -8 \pi G \rho_b \left[ D (E-D^2) 
\left (\psi^{(1)}_{,ii} \psi^{(2)}_{,jj}
 -\psi^{(1)}_{,ij} \psi^{(2)}_{,ji} \right )
 + D^3 \mbox{det} (\psi^{(1)}_{,ij}) \right ]  \,.
\label{eqn:Lagrange-3rd}
\end{eqnarray}

The spacial component of the second order are written
as follows:
\begin{equation}
\psi^{(2)}_{,ii} = (\psi_{,jj}^{(1)})^2
 - \psi_{,jk}^{(1)} \psi_{,kj}^{(1)} \,.
\end{equation}
The time component of second-order $E$ obeys the evolution equation.
\begin{equation}
\ddot{E} + 2 \frac{\dot{a}}{a} \dot{E} - 4 \pi G \rho_b E
= -2 \pi G \rho_b D^2 \,.
\end{equation}
For the third order, we divide to two components of which one is
derived from combining $\psi^{(1)}$ and $\psi^{(2)}$ and the other
is derived from $(\psi^{(1)})^3$.
\begin{eqnarray}
S^{(3)} &=& F_a (t) \psi_a^{(3)} (\bm{q}) + F_b (t) \psi_b^{(3)} (\bm{q})
\,, \\
\psi^{(3)}_{a~,ii} &=& \mbox{det} \left (\psi^{(1)}_{,jk} \right ) \,, \\
\psi^{(3)}_{b~,ii} &=& \psi^{(1)}_{,jj} \psi^{(2)}_{,kk}
 -\psi^{(1)}_{,jk} \psi^{(2)}_{,kj} \,, \\
\ddot{F}_a + 2 \frac{\dot{a}}{a} \dot{F}_a - 4 \pi G \rho_b F_a
&=& -8 \pi G \rho_b D^3 \,, \\
\ddot{F}_b + 2 \frac{\dot{a}}{a} \dot{F}_b - 4 \pi G \rho_b F_b
&=& -8 \pi G \rho_b D (E-D^2) \,.
\end{eqnarray}
For the E-dS Universe model, $E$ and $F$ and described with
analytic form.%
\begin{eqnarray}
E(t) &=& -\frac{3}{7} t^{4/3} \propto D(t)^2 \,, \\
F_a(t) &=& -\frac{1}{3} t^2 \propto D(t)^3 \,, \\
F_b(t) &=& \frac{10}{21} t^2 \propto D(t)^3 \,.
\end{eqnarray}
For the non-flat Universe model, $E$ can be described with
analytic form. For the open Universe model,
\begin{equation}
E(\tau) = -\frac{1}{2} - \frac{9}{2} (\tau^2-1)
 \left [ 1+\frac{\tau}{2} \log \left (\frac{\tau-1}{\tau+1}
 \right ) + \frac{1}{2} \left (\tau + \frac{\tau^2-1}{2}
 \log \left (\frac{\tau-1}{\tau+1} \right ) \right ) \right ] \,,
\end{equation}
where $\tau$ is defined by Eq.~(\ref{eqn:def-newtime}).
For the closed model ($\Omega >1$), we obtain analytic form
by the transformation $\tau \rightarrow i \tau$ as
in Eq.~(\ref{eqn:LA-open-1st}).

For the $\Lambda$-flat model, unfortunately, we need to solve
evolution equations with the numerical method~\cite{Bouchet95}.
\begin{eqnarray}
3 (h^2-1) \ddot{E} + 2h \dot{E} &=& 2E - 2D^2 \,, \\
3 (h^2-1) \ddot{F}_a + 2h \dot{F}_a &=& 2 F_a - 2 D^3 \,, \\
3 (h^2-1) \ddot{F}_b + 2h \dot{F}_b &=& 2 F_b + 2 D^3
 \left (1- \frac{E}{D^2} \right ) \,.
\end{eqnarray}

\subsection{The transverse mode}\label{subsec:Trans}

For the transverse mode, until third-order solutions have been obtained
\cite{Buchert92,Barrow93,Sasaki98}. The evolution equation
for the first-order perturbation is given as
\begin{equation}
\ddot{\bm{S}}^T + 2 \frac{\dot{a}}{a} \dot{\bm{S}}^T = 0 \,.
\end{equation}
For the E-dS model, the first-order
solution is written as
\begin{equation}
\bm{S}^T ({\bm q}, t) = \bm{S}_a^T ({\bm q})
 + t^{-1/3} \bm{S}_b^T ({\bm q})\,.
\end{equation}
Therefore the transverse mode does not have a growing solution
in the first-order perturbation. Furthermore, if we consider only
the longitudinal mode for the first order, the second-order
transverse mode
does not appear. However, the third-order transverse mode is generated
by the triplet of the first-order longitudinal mode, and it grows.
Although the third-order solution has the growing mode, this solution does
not show that a vorticity grows.

\subsection{Modified models I -- avoidance of formation of a caustic}
\label{subsec:modify1}

Cosmological N-body simulations show that pancakes,
skeletons, and clumps remain during evolution. However, when we
continue applying the solutions of ZA, PZA, or PPZA after the appearance
of caustics, the nonlinear structure diffuses and breaks.

Adhesion approximation (AA)~\cite{Gurbatov89} was proposed from
a consideration based on Burgers' equation.
This model is derived by the addition of an artificial viscous term to ZA.
AA with small viscosity deals with the
skeleton of the structure, which at an arbitrary time is found directly
without a long numerical calculation.

We briefly describe the adhesion model. In ZA, the equation for
``peculiar velocity'' in the E-dS model is written as follows:
\begin{eqnarray}
\frac{\partial \bm{u}}{\partial a} + (\bm{u} \cdot \nabla_x) \bm{u}
&=& 0 \,, \\
\bm{u} \equiv \frac{\partial \bm{x}}{\partial a}
 = \frac{\dot{\bm{x}}}{\dot{a}} \,,
\end{eqnarray}
where $a (\propto t^{2/3})$ is scale factor.
To go beyond ZA, we add the artificial viscosity term to the right side
of the equation.
\begin{equation} \label{eqn:adhesion}
\frac{\partial \bm{u}}{\partial a} + (\bm{u} \cdot \nabla_x) \bm{u}
= \nu \nabla_x^2 \bm{u} \,.
\end{equation}

We consider the case when the viscosity coefficient $\nu \rightarrow +0$
 ($\nu \ne 0$). In this case, the viscosity term especially affects the
high-density region. Within the limits of a small $\nu$, the analytic solution of
Eq.(\ref{eqn:adhesion}) is given by
\begin{equation}
\bm{u} (\bm{x}, t) = \sum_{\alpha} \left(\frac{\bm{x}-\bm{q}_{\alpha}}
{a} \right ) j_{\alpha} \exp \left( -\frac{I_{\alpha}}{2\nu} \right )
/ \sum_{\alpha} j_{\alpha} \exp \left( -\frac{I_{\alpha}}{2\nu} \right )
\,,
\end{equation}
where $\bm{q}_{\alpha}$ is the Lagrangian points that minimize the
action
\begin{eqnarray}
I_{\alpha} & \equiv & I(\bm{x}, a; \bm{q}_{\alpha})
 = S_0 (\bm{q}_{\alpha}) + \frac{(\bm{x}-\bm{q}_{\alpha})^2}{2a}
 = \mbox{min.} \,, \\
j_{\alpha} & \equiv & \left. \left[ \det \left (\delta_{ij}
 + \frac{\partial^2 S_0}{\partial q_i \partial q_j} \right )
 \right ]^{-1/2} \right |_{\bm{q}=\bm{q}_{\alpha}} \,, \\
S_0 &=& S(\bm{q}, t_0) \,,
\end{eqnarray}
considered as a function of $\bm{q}$ for fixed $\bm{x}$~\cite{Kofman92}.
In AA, because of the viscosity term, the caustic does not appear and
a stable nonlinear structure can exist. However, when the evolution is
advanced too much, the Universe becomes covered with a high-density structure
called  ``skeleton.'' 

As another modification theory, truncated (or Optimized)
Zel'dovich approximation (TZA)
has been used well~\cite{Coles93, Melott94B}.
During
evolution, the small scale structure contracts and forms caustics. Therefore,
if we introduce some cutoff in the small scale, we will be able to avoid 
the formation of caustics~\cite{Coles93, Melott94B}.

In TZA, for the avoidance
of caustics, we introduce a cutoff in the initial density spectrum.
Various cutoff methods were considered, and the Gaussian cutoff
was the most suitable from the comparison with N-body simulation.
The Gaussian cutoff is introduced to the initial density
spectrum as follows:
\begin{equation}
{\cal P}(k,t_{\rm in}) \rightarrow {\cal P}(k, t_{\rm in}) \exp
\left (-k^2/k_{NL} \right ) \,,
\end{equation}
where $k_{NL}$ is the ``nonlinear wavenumber'', defined by
\begin{equation} \label{eqn:TZA-knl}
1 = D_+(t)^2 \int_{k_0}^{k_{NL}} {\cal P}(k,t_{\rm in}) dk \,,
\end{equation}
where $D(t)_+$ is the growing factor in ZA (Eq.~(\ref{eqn:sol-ZA})).
Because the ``nonlinear wavenumber'' depends on the time, the
wavenumber becomes small during evolution. Therefore,
when the evolution is advanced too much,
small structure will vanished.

\subsection{Modified models II -- high-accuracy models}
\label{subsec:modify2}
Bagla and Padmanabhan~\cite{Bagla94} developed a new approximation
scheme that improved ZA. They considered extension the frozen
flow approximation~\cite{Matarrese92}. Then they showed better
agreement with the N-body simulations than ZA.

Yoshisato \textit{et al.}~\cite{Yoshisato98} introduced the 
Pad\'{e} approximation for the evolution of the density fluctuation.
The Pad\'{e} approximation seems to be a generalization of the Taylor
expansion. For a given function $f(x)$, the Pad\'{e} approximation
is written as the ratio of two polynomials:
\begin{equation}
f(x) \simeq \frac{\sum_{k=0}^M a_k x^k}{1+ \sum_{k=1}^N b_k x^k} \,,
\end{equation}
where $a_k$ and $b_k$ are constant coefficients. Assume we already know
the coefficient $c_l$ ($0 \le l \le M+N$)
of the Taylor expansion around $x=0$.
\begin{equation}
f(x) = \sum_{l=0}^{M+N} c_l x^l + o(x^{M+N+1}) \,.
\end{equation}
Comparing the coefficents $a_k$, $b_k$, and $c_k$, we determine
$a_k$ and $b_k$.
\begin{eqnarray}
a_0 &=& c_0 \,, \\
a_k &=& \sum_{m=1}^N b_m c_{k-m} ~~(k=1,\cdots,N) \,, \\
\sum_{m=1}^N b_m c_{N-m+k} &=& -c_{N+k} ~~(k=1,\cdots,N) \,.
\end{eqnarray}
The advantage of the Pad\'{e} approximation is that
even if we consider a same-order expansion,
the Pad\'{e} approximation describes original function
rather better than the Taylor expansion.

First, Yoshisato \textit{et al.}~\cite{Yoshisato98} showed that
the application of the Pad\'{e} approximation for Eulerian description
improves the past Eulerian models. Then Matsubara
\textit{et al.}~\cite{Matsubara98} proposed the application of
the Pad\'{e} approximation for Lagrangian description and
improved PZA and PPZA.

Another approach is known as ``local'' approximations. In ZA,
the growing
factor $D$ is independent of the position. On the other hand,
in local approximations, the factor depends on position.

First, we derive an evolution equation for the eigenvalue
of the deformation tensor (Eq.~(\ref{eqn:ZA-deformation})).
Here we use time variable $\tau$ (Eq.~(\ref{eqn:def-newtime})).
Differentiating of Eq.~(\ref{eqn:Lagrangian-Euler}) with
respect to $x_j$ and summing up the trace part,
we obtain
\begin{equation} \label{eqn:Raychaudhuri}
X_{ij}^{-1} \frac{{\rm d}^2 X_{ji}}{{\rm d} \tau^2}
= -4 \pi G a^4 \rho_b (J^{-1} -1) \,.
\end{equation}
If $X_{ij}$ is diagonal,
\begin{equation}
X_{ij} = (1+w_i) \delta_{ij} \,,
\end{equation}
Eq.~(\ref{eqn:Raychaudhuri}) is written as
\begin{equation} \label{eqn:Raychaudhuri2}
\sum_{i=1}^3 \frac{\ddot{w}_i}{(1+w_i)}
= -4 \pi G a^4 \rho_b
 \left [ \frac{1}{(1+w_1)(1+w_2)(1+w_3)} -1 \right ] \,.
\end{equation}
For a spherical case we have $w_1=w_2=w_3$;
for a cylindrical case $w_1=w_2$ and $w_3=0$;
and for a planar case $w_2=w_3=0$.
ZA is a linearized form for $w_i$ of Eq.~(\ref{eqn:Raychaudhuri2}).

Reiseneger and Miralda-Escud\'{e}~\cite{Reisenegger95} 
proposed changing the eigenvalue
(Eq.~(\ref{eqn:ZA-eigenvalue})) as
\begin{equation}
w_i = -D(t, \lambda_i^0) \lambda_i^0 (\bm{q}) \,.
\end{equation} 
This is then substituted in Eq.~(\ref{eqn:Raychaudhuri2}).
They named this model the modified Zel'dovich
approximation (MZA) or extension of ZA (EZA). The MZA is
exact for planar, spherical, and cylindrical symmetric cases.
However, they also reported that the MZA may not work
for underdense regions.

Audit and Alimi~\cite{Audit96} introduced another ansatz.
Eq.~(\ref{eqn:Raychaudhuri2}) may be written in this form:
\begin{equation} \label{eqn:Raychaudhuri3}
\sum_{i=1}^3 \left [ (1+w_j + w_k + w_j w_k) \ddot{w_i}
 - 4 \pi G a^4 \rho_b \left (1+\frac{w_j+w_k}{2}
 + \frac{w_j w_k}{3} \right ) w_i \right ] =0 \,.
\end{equation}
They split this equation into three equations for each $w_i$:
\begin{equation} \label{eqn:DTA}
(1+w_j + w_k + w_j w_k) \ddot{w_i}
 = 4 \pi G a^4 \rho_b \left (1+\frac{w_j+w_k}{2}
 + \frac{w_j w_k}{3} \right ) w_i \,.
\end{equation}
This approximation is called the Deformation Tensor Approximation
(DTA)~\cite{Audit96}. Although this is also exact for
planar, spherical, and cylindrical symmetric cases,
the splitting of Eq.~(\ref{eqn:Raychaudhuri3}) is not unique,
and we could add more local terms in Eq.~(\ref{eqn:DTA}).

Betancort-Rijo and L\'{o}pez-Corredoida~\cite{Betancort-Rijo00}
proposed another generalization. In terms of the linear
solution $\lambda_i = D(t) \lambda^0_i (\bm{q})$ in ZA,
they generalized Eq.~(\ref{eqn:ZA-eigenvalue}) to
\begin{equation} \label{eqn:CZA-lambda}
w_i = - r_i(\lambda_i, \lambda_j, \lambda_k) \lambda^0_i (\bm{q}) \,,
\end{equation}
where $r_i(\lambda_i, \lambda_j, \lambda_k)$ is the power series
of $\lambda$.
\begin{equation}
r_i(\lambda_i, \lambda_j, \lambda_k)
= 1+ \sum_{l, m, n=0}^{\infty} C_{l, m, n}^p
 (\lambda_j + \lambda_k)^l (\lambda_j - \lambda_k)^{2n}
 \lambda_i^m \,,
\end{equation}
where $C_{l, m, n}^p$ are the coefficients of the $p$-th order
terms ($p=l+2n+m$). The ZA corresponds to $r_i=1$.
They named this generalized model the
``Complete Zel'dovich approximation''
(CZA).
They calculated explicitly the coefficients $C_{l, m, n}^p$
up to the fourth order of $\lambda$ in the E-dS Universe model
~\cite{Betancort-Rijo00}.
This model is also exact for
planar, spherical, and cylindrical symmetric cases. Furthermore,
the CZA describes the evolution of ellipsoid dust better than
the MZA, the DTA, and ZA~\cite{Makler01}.
However the CZA does not apply for perturbations with negative
values of $\lambda_i^0$~\cite{Betancort-Rijo00, Makler01}.

\section{Pressure model}\label{sec:pressure}
Although AA seems a good model for avoiding the formation of caustics,
the origin of the modification (or artificial viscosity) is not
clarified. Buchert and Dom\'{\i}nguez~\cite{Buchert98} argued that the effect
of velocity dispersion becomes important beyond the caustics.
They showed that when the velocity dispersion is still
small and can be considered isotropic, it gives effective
``pressure'' or viscosity terms.
Buchert \textit{et al.}~\cite{Buchert99} showed how the viscosity term
is generated by the effective pressure of a fluid
under the assumption that the peculiar acceleration is
parallel to the peculiar velocity.

Adler and Buchert~\cite{Adler99} have
formulated the Lagrangian perturbation theory for a barotropic fluid.
Morita and Tatekawa~\cite{Morita01} and Tatekawa \textit{et al.}~\cite{Tatekawa02}
solved the Lagrangian perturbation equations for a polytropic fluid
in the Friedmann Universe.
Hereafter, we call this model the ``pressure model''.

If the pressure is given by a barotropic equation of state,
the pressure affects only the longitudinal mode in the first-order.
Therefore, the transverse solution in the pressure model
becomes the same as that in the dust model.
The evolution equations are written as
\begin{eqnarray}
\ddot{\bm{S}}^T + 2 \frac{\dot{a}}{a} \dot{\bm{S}}^T &=& 0 \,,
\label{eqn:P-trans-1st} \\
\ddot{S} + 2 \frac{\dot{a}}{a} \dot{S} - 4 \pi G \rho_b S
- \frac{1}{a^2} \left. \frac{{\rm d} P}{{\rm d} \rho}
 \right |_{\rho=\rho_b} \nabla^2 S &=& 0 \,.
\label{eqn:P-longi-1st}
\end{eqnarray}

When we assume the polytropic equation of state $P=\kappa \rho^{\gamma}$,
the first-order equation for the longitudinal mode is written as follows.
\begin{equation} \label{eqn:P-1stL}
\ddot{S} + 2\frac{\dot{a}}{a} \dot{S} - 4 \pi G \rho_b S
- \frac{\kappa \gamma \rho_b^{\gamma-1}}{a^2}
 \nabla^2 S = 0 \,.
\end{equation}
The solution manifestly depends on the scale of the fluctuation. We
adopt Fourier transformation in Lagrangian space for Eq.~(\ref{eqn:P-1stL}).
\begin{equation} \label{eqn:P-1stLF}
\ddot{\widehat{S}} + 2\frac{\dot{a}}{a} \dot{\widehat{S}}
 - 4 \pi G \rho_b \widehat{S}
 + \frac{\kappa \gamma \rho_b^{\gamma-1}}{a^2} \left | \bm{K} \right |^2
 \widehat{S} = 0 \,,
\end{equation}
where $\bm{K}$ is a Lagrangian wavenumber. 
In the E-dS Universe model, we can write the first-order solution
relatively simply~\cite{Morita01}. For $\gamma \ne 4/3$,
\begin{equation}\label{eqn:hatSbessel}
\widehat{S}(\bm{K},a) \propto a^{-1/4}
\, \mathcal{J}_{\pm 5/(8-6\gamma)}
\left( \sqrt{\frac{2C_2}{C_1}}
\frac{|\bm{K}|}{|4-3\gamma|}
\, a^{(4-3\gamma)/2} \right) \,,
\end{equation}
where $\mathcal{J}_{\nu}$ denotes the Bessel function of order $\nu$,
and for $\gamma=4/3$,
\begin{equation}\label{eqn:hatS43}
\widehat{S}(\bm{K},a) \propto
a^{-1/4 \pm \sqrt{25/16 - C_2 |\bm{K}|^2 / 2C_1}} \,,
\end{equation}
where $C_1 \equiv 4 \pi G \rho_{\rm b}(a_{\rm in})
\, a_{\rm in}^{\ 3} /3$
and $C_2 \equiv \kappa \gamma \rho_{\rm b}(a_{\rm in})^{\gamma-1}
\, a_{\rm in}^{\ 3(\gamma-1)}$. $a_{\rm in}$
is scale factor when an initial condition is given.

For the other FRW universe model, if $\gamma$ takes a special value,
we obtain an analytic solution with a hypergeometric
function~\cite{Tatekawa02}: For
open or closed model, $\gamma=1, 4/3$. For the $\Lambda$-flat
Universe model, $\gamma= 1/3, 4/3$.

In this model, the behavior of the solutions strongly depends on
the relation between the scale of fluctuation and the Jeans scale.
Here we define the Jeans wavenumber as
\begin{equation} \label{eqn:Pressure-Jeans}
K_{\rm J} \equiv \left(
\frac{4\pi G\rho_{\rm b} a^2}
     {{\rm d} P / {\rm d} \rho (\rho_{\rm b})} \right)^{1/2} \,.
\end{equation}
The Jeans wavenumber, which gives a criterion for
whether a density perturbation with a wavenumber
will grow or decay with oscillation,
depends on time in general. If the polytropic index $\gamma$ is smaller
than $4/3$, all modes become decaying modes and the fluctuation
will disappear. On the other hand, if $\gamma > 4/3$, all density
perturbations will grow to collapse. In the case where $\gamma=4/3$,
the growing and decaying modes coexist at all times.

We rewrite the first-order solution Eq.~(\ref{eqn:hatSbessel}) with
the Jeans wavenumber:
\begin{equation}
\widehat{S}(\bm{K},a) \propto a^{-1/4}
\, \mathcal{J}_{\pm 5/(8-6\gamma)}
\left( \frac{\sqrt{6}}{|4-3\gamma|}
\frac{|\bm{K}|}{K_{\rm J}} \right) \,.
\end{equation}

In the pressure model, the second-order perturbative solutions have
been derived~\cite{Morita01, Tatekawa02}. Because of mode coupling,
the calculation of the second-order perturbative solutions are
complicated. Only for the E-dS Universe model and the case where
$\gamma=4/3$ have the solutions of the longitudinal and the transverse
mode been derived~\cite{Morita01}. Then, in the E-dS Universe model,
if $\gamma$ is larger than $4/3$, it was shown that the effect
of the second-order pressure becomes weak gradually during
evolution~\cite{Tatekawa02}. In this paper, the authors considered
the expansion of the Bessel function,
\begin{equation}
{\mathcal J}_{\nu} (z) = \left ( \frac{z}{2} \right )^{\nu}
 \sum_{n=0}^{\infty} \frac{(-1)^n (z/2)^{2n}}{n! \Gamma(\nu+n+1)} \,.
\end{equation}
In the E-dS Universe model, $z \propto a^{(4-3\gamma)/2}$
(Eq.~(\ref{eqn:hatSbessel})).
If $\gamma$ is larger than $4/3$, the index of argument of the Bessel function
becomes negative. Therefore, precision does not become much worse even if we
drop the higher terms of the expansion. Using finite expansion, they analyzed
the second-order perturbation for the case of $\gamma=5/3$.

These models are considered for a single component fluid. Of course,
there are several components in the Universe, for example, baryonic matter
and dark matter. Matarrese and Mohayaee~\cite{Matarrese02} proposed
the Lagrangian perturbation model for two component fluids. One is
dark matter which affects only gravity. Another is baryonic matter
which affects gravity for all fluid and pressure for baryonic matter.

When we consider more than one species of matter,
the evolution equation of the density fluctuation is
modified~\cite{Paddy93}.
We assume a polytropic equation of state for baryonic matter.
\begin{equation}
P \propto \rho^{\gamma} \,.
\end{equation}

Taking the divergence of the baryon Euler equation,
we obtain the evolution equation
for the baryon density fluctuation with the Lagrangian
perturbation.
\begin{equation} \label{eqn:delta-baryon}
\nabla_x \cdot \bm{s}_B'' =
-\frac{3}{2a} \left [ \nabla_x \cdot \bm{s}_B'
 + \frac{\delta_{DM}}{a} + \frac{1}{(\gamma-1) a k_J^2}
 \nabla_x^2 (1+\delta_B)^{\gamma-1} \right ] \,
,\end{equation}
where $B$ and $DM$ mean baryon and dark matter, respectively.
$k_J$ is the Jeans wavenumber of the baryonic matter.
$(')$ denotes the derivative by a scale factor ($' = \partial/\partial a$).
Here we assume that the amount of the baryonic matter is
much less than that of the dark matter. Therefore, we ignore
the gravitational force of the baryonic matter.
Substituting Eq.~(\ref{eqn:L-Jacobian}) for (\ref{eqn:delta-baryon}),
we obtain
\begin{equation}
\nabla_x \cdot \bm{s}_B'' + \frac{3}{2a} \nabla_x \cdot
 \bm{s}_B' = -\frac{3}{2 a^2} \frac{1-J_{DM}}{J_{DM}}
 - \frac{3}{2(\gamma-1) a^2 k_J^2}
 \nabla^2 \left (\frac{1-J_B}{J_B} \right )^{\gamma-1} \,.
\end{equation}
Next, we consider only first-order terms for the Lagrangian
displacement $\bm{s}$.
\begin{equation}
\nabla_q \cdot \bm{s}_B'' + \frac{3}{2a} \nabla_q \cdot
 \bm{s}_B' -\frac{3}{2 a^2} \nabla_q \cdot \bm{s}_{DM}
= \frac{3}{2a^2 k_J^2} \nabla_q^2 \nabla_q \cdot \bm{s}_B
 \,.
\end{equation}
We note that the same Eulerian position $\bm{x} (t)$ is generally
reached by the two components from different Lagrangian
positions $\bm{q}_B$ and $\bm{q}_{DM}$. Here we ignore this
difference and set simply $\bm{q}_B = \bm{q}_{DM}$, because
we consider the lowest order approximation.

We assume that the perturbation has only a longitudinal mode
($\bm{s} = \nabla_q S$). Under this assumption, we obtain
the evolution equation of the Lagrangian perturbation for
the baryonic matter.
\begin{equation}
S_B'' + \frac{3}{2a} S_B' - \frac{3}{2a^2} \frac{1}{k_J^2}
\nabla_q^2 S_B = \frac{3}{2a} S_{DM} \,.
\end{equation}
This equation can be solved in Lagrangian Fourier space.

\section{Summary}\label{sec:summary}

We briefly review Lagrangian perturbation theory
in Newtonian cosmology. From a comparison to Eulerian
perturbation theory, Lagrangian perturbation theory
describes quasi-nonlinear evolution of density fluctuation
rather well. We notice that when we analyze physical phenomena
or statistical quantities with Lagrangian perturbation,
because the physical coordinates are Eulerian ones,
we must transform the coordinates from Lagrangian ones to
Eulerian ones. Although transformation from Eulerian ones
to Lagrangian ones is given by Eq.~(\ref{eqn:Euler-Lagrange}),
the inverse transformation seems a little difficult.

Although Lagrangian perturbation theory seems to work well for
describing quasi-nonlinear evolution of the density
fluctuation, it is not always useful. For example, we consider
the evolution of the spherical void~\cite{Munshi94,Sahni96}.
In general, when we consider higher-order perturbation,
the approximation is improved. However, for void evolution,
when we advance for a long time, higher-order terms
show bad behavior. The even-order (2nd, 4th,...) perturbative
terms promote to development of the reverse, i.e., the void
disappears and the density fluctuation converges.
On the other hand, the odd-order (3rd, 5th,...) perturbative
terms promote evolution excessively. We provie details
in Appendix~\ref{sec:void}.

In Sec.~\ref{subsec:modify1}, we noticed modified models
for avoidance of formation of a caustic. However, the origin
of those modifications is not clarified. 
Buchert \textit{et al.}~\cite{Buchert99} discussed the relation
between the viscosity term in the Adhesion approximation
and the effective pressure of a fluid.
Can the pressure become the origin of the modification?
According to a recent study~\cite{Tatekawa04B}, the pressure term
can realize that the density distribution looks like that of AA
until reaching quasi-nonlinear regime. However, because of
Jeans instability, the behavior of the density fluctuation in
the pressure model seems different to that in AA.

On the other hand, the pressure especially affects small-scale
structure. Therefore, we can expect that the pressure term
realize the Gaussian cutoff in TZA. In fact, the correspondence
between ``nonlinear wavenumber'' $k_{NL}$
in TZA (Eq.~(\ref{eqn:TZA-knl}))
and the Jeans wavenumber $K_J$ in the pressure model
(Eq.~(\ref{eqn:Pressure-Jeans})) has been
discussed~\cite{Tatekawa04A}.
As a result, the correspondence has not been explained 
sufficiently.
The reason is as follows:
First, in TZA, $k_{NL}$ affects only the initial spectrum.
On the other hand,
$K_J$ affects the evolution of fluctuation. Second, although
$k_{NL}$ obviously depends on the initial spectrum, we did not clarify
the dependence on the initial
condition of $K_J$. We think that a consideration of the physical
process, which was not considered here, or the analysis of
the N-body simulation, is necessary for a decision of $K_J$,
i.e. $\kappa$.

Therefore, the effect of pressure cannot explain modifications
in the Lagrangian approximation. If we consider the origin of
these modification with the effect of velocity dispersion,
we may have to consider anisotropic velocity dispersion
~\cite{Maartens99}.

The Lagrangian perturbation theory seems rather useful until
quasi-nonlinear regime develops. However,
the density fluctuation becomes
strongly nonlinear in the present Universe. Can we apply
the theory to the problem of structure formation?
Because a huge simulation has executed (\cite{Evrard02}),
someone may thinks that the perturbation theory is no longer useful.
We think that analytic methods are important
for examining the physical process of structure formation.
For example, in the discussion about statistical quantities,
the analytical method shows remarkable results
~\cite{Davis77,Nityananda94,Paddy96,Yano97}.

One possibility is analysis of the past structure.
Recently, several galaxy redshift survey projects have been
progressing~\cite{Hawkins03,Abazajian04,Gerke04}. In these projects,
many galaxies within a region ($z<0.3$ for 2dF,
$z<0.5$ for SDSS, and $0.7 < z < 1.4$ for DEEP2) have
been observed. For the next generation of spectroscopic survey,
for example, Kilo-Aperture Optical Spectrograph (KAOS) has been
proposed~\cite{KAOS}. The two primary scientific purposes of KAOS are
(i) the determination of the equation of state of dark energy
and (ii) the study of the origin of galaxies. As one of the other
science applications for KAOS, the growth of structure is
considered. In this project, many high-$z$ galaxies will be observed.
Therefore, this project would not only probe topology of
the matter distribution but also the dynamical history of
structure formation.
Because the density contrast will still be small in
the high-z region, we expect that we will be able to discuss
the characters of the density
fluctuation using the Lagrangian perturbation theory well.

Recently, various dark matter models have been
proposed~\cite{Ostriker03}.
Some of them affect not only the gravity but also a special interaction.
We also show that the linear approximation of the pressure model
seems rather good until a quasi-linear regime develops.
If the interactions of the dark matter are affected by effective pressure,
the linear approximation can be applied for the analysis
of the quasi-nonlinear evolution of the density fluctuation.

\begin{acknowledgments}
We are grateful to Kei-ichi Maeda for his continuous encouragement.
We thank Shuntaro Mizuno and Thanu Padmanabhan
for useful discussion.
We also thank Kenta Kiuchi and Masato Nozawa for
manuscript corrections.
We would like to thank Peter Musolf for checking of
English writing of this paper. 
This work was supported by a Grant-in-Aid for Scientific
Research Fund of the Ministry of Education, Culture, Sports, Science
and Technology (Young Scientists (B) 16740152).
\end{acknowledgments}

\appendix
\section{Spherical void evolution with dust fluid} \label{sec:void}

We describe the development of the spherical void with Lagrangian
perturbation. Here we consider ``top-hat'' spherical void, i.e.,
a constant density spherical void. In the E-dS Universe model,
the equation of motion of a spherical shell is written as
\begin{equation} \label{eqn:spherical-void}
\frac{\rm{d}}{{\rm d} t} \left (a^2 \frac{{\rm d} x}{{\rm d} t} \right )
= -\frac{2a^2 x}{9t^2} \left [ \left (\frac{x_0}{x}\right )^3 -1
 \right ] \,,
\end{equation}
where $x$ is a comoving radial coordinate and $x_0=x(t_0)$~\cite{Munshi94}.
Under the initial condition $\delta=a$ for $a \rightarrow 0$,
Eq.~(\ref{eqn:spherical-void}) can be integrated.
\begin{equation} \label{eqn:spherical-void2}
\left (\frac{{\rm d} R}{{\rm d} a} \right )^2 = a
 \left ( \frac{1}{R} - \frac{3}{5} \right ) \,,
\end{equation}
where $R(\theta) =a(t) x/x_0$ is physical particle trajectory.
The exact solution for the expansion of a top-hat void
(Eq.~(\ref{eqn:spherical-void2})) can be parameterized as follows:
\begin{eqnarray}
R(\theta) &=& \frac{3}{10} \left (\cosh \theta -1 \right ) \,, \\
a(\theta) &=& \frac{3}{5} \left [ \frac{3}{4}
 \left (\sinh \theta - \theta \right ) \right ]^{2/3} \,.
\end{eqnarray}
From these equations, we can obtain density fluctuation.
\begin{eqnarray}
\delta (x) &=& \left ( \frac{x_0}{x} \right )^3 -1 \nonumber \\
 &=& \frac{9 (\theta-\sinh \theta)^2}{2 (\cosh \theta -1)^3}
 -1 \,.
\end{eqnarray}
The perturbative solution for the void
in the E-dS Universe model is given by
\begin{equation} \label{eqn:Lagrange-void}
R(t)=R_0 \left [1- \sum_{k=1}^n (-1)^k C_k a^k \right ] \,,
\end{equation}
where $C_k$ is given by Lagrangian perturbative equations.
Munshi, Sahni, and Starobinsky~\cite{Munshi94} derived
up to the third-order perturbative solution ($C_1, C_2, C_3$).
In addition to these,
Sahni and Shandarin~\cite{Sahni96} obtained $C_4$ and $C_5$.
They concluded that ZA remains the best approximation to
apply to the late-time evolution of voids. Especially,
if we stop to improve until even order (2nd, 4th, 6th, $\cdots$),
the perturbative solution describes the contraction of void.
From the viewpoint
of the convergence of the series, the conclusion seems
reasonable. Table~\ref{tab:Void} shows the coefficients
of Lagrangian perturbation theory for top-hat spherical voids.
Here we derived up to the eleventh order. The convergence radius of
the series $a_c$ is defined by the ratio of the coefficients:
\begin{equation}
a_c \equiv \lim_{k \rightarrow \infty} \frac{a_k}{a_{k+1}} \,.
\end{equation}
When the scale factor $a$ becomes larger than $a_c$, the
series (Eq.~(\ref{eqn:Lagrange-void})) does not converge.
\begin{table}
\caption{\label{tab:Void}
The ratio of coefficients of Lagrangian perturbation theory
for top-hat spherical voids ($C_{k-1}/C_k$).
The higher order the perturbation we consider, the worse
the approximation becomes in late-time evolution.
In other words,
when the order of the perturbation increases, the convergence of the
series becomes bad.}
\vspace{0.3cm}
\begin{ruledtabular}
\begin{tabular}{lccccc}
$k$ & $2$\footnotemark[1]
 & $3$\footnotemark[1]
 & $4$\footnotemark[2]
 & $5$\footnotemark[2]
 & $6$  \\ \hline
$C_{k-1}/C_k$ & $7$
 & $3.52174$ & $2.80517$
 & $2.49236$ & $2.31645$ \\ \hline
\end{tabular}
\vspace{0.3cm}
\begin{tabular}{lccccc}
$k$ & $7$ & $8$ & $9$ & $10$ & $11$ \\ \hline
$C_{k-1}/C_k$ 
 & $2.20357$ & $2.12495$ & $2.06704$
 & $2.02260$ & $1.98743$
\end{tabular}
\end{ruledtabular}
\footnotetext[1]{Munshi, Sahni, and Starobinsky~\cite{Munshi94}}
\footnotetext[2]{Sahni and Shandarin~\cite{Sahni96}}
\end{table}

We estimate the convergence radius from finite series.
Table~\ref{tab:Void} shows the ratio $C_{k-1}/C_k$.
When $k$ is increased, $C_{k-1}/C_k$ decreases. Therefore,
the higher order the perturbation we consider, the worse
the approximation becomes in late-time evolution.
If we consider only normal Lagrangian perturbation,
the problem cannot be solved. The Pad\'{e} approximation
\cite{Yoshisato98,Matsubara98} shows that the approximation
for the late-time evolution of voids is improved, and
therefore some applications are expected.


\section{Quantities used in this paper} \label{sec:quantities}
In this paper, we use the following variables for physical quantities,
coordinates, and so on.
\vspace{-1cm}
\begin{table}[H]
\caption{\label{tab:coor-variable} Coordinate and velocity variables.
}
\begin{tabular}{c|l|c|l} \hline \hline
$\bm{r}$     & physical coordinates &
$\bm{x}$  & comoving Eulerian coordinates \\
$\bm{q}$     & Lagrangian coordinates &
$\bm{u}$  & velocity in physical coordinates  \\
$\bm{v}$     & peculiar velocity && \\ \hline
\end{tabular}
\end{table}
\vspace{-3cm}
\begin{table}[H]
\caption{\label{tab:time-variable} Time variables.
}
\begin{tabular}{c|l|c|l} \hline \hline
$t$ & standard cosmological time &
$\tau$     & new time variable (Eq.~(\ref{eqn:def-newtime})) \\
$a$     & scale factor &
$z$       & red shift \\ \hline \hline
\end{tabular}
\end{table}
\vspace{-2cm}
\begin{table}[H]
\caption{\label{tab:cosmo-variable} Cosmological parameters.
}
\begin{tabular}{c|l|c|l} \hline \hline
$H=\dot{a}/a$ & Hubble parameter &
$\mathcal{K}$ & curvature constant \\
$\Lambda$ & cosmological constant &
$\Omega$ & density parameter \\ \hline
\end{tabular}
\end{table}
\vspace{-2cm}
\begin{table}[H]
\caption{\label{tab:phys-variable} Physical quantities of matter.
}
\begin{tabular}{c|l} \hline \hline
$\rho$   & matter density \\
$\rho_b$    & background (averaged) matter density \\
$P$      & pressure of matter \\
$G$         & gravitational constant  \\
$\Phi$   & gravitational potential \\
$\tilde{\bm{g}}$ & gravitational force in comoving coordinates \\
$\delta$ & density fluctuation (Eq.~(\ref{eqn:def-delta})) \\
$\bm{s}$    & Lagrangian displacement (perturbation) vector \\
$\bm{\omega}$ & vorticity (Eq.~(\ref{eqn:vorticity})) \\ \hline
\end{tabular}
\end{table}

\begin{table}[H]
\caption{\label{tab:S-variable} Lagrangian perturbation.
}
\begin{tabular}{c|l} \hline \hline
$S$         & Lagrangian perturbation potential (longitudinal mode) \\
$\bm{s}^T$ & Lagrangian perturbation (transverse mode) \\
$J$        & Jacobian of the coordinate transformation from
 $\bm{x}$ to $\bm{q}$ \\
$\psi$      & spacial part of Lagrangian perturbation potential \\
$D_+$      & growing factor in ZA (Eq.~(\ref{eqn:sol-ZA})) \\
$D_-$    & decaying factor in ZA  (Eq.~(\ref{eqn:sol-ZA})) \\
$E$        & growing factor in PZA (Eq.~(\ref{eqn:Lagrange-2ndL})) \\
$F$     & growing factor in PPZA (Eq.~(\ref{eqn:Lagrange-3rd})) \\
$X_{\alpha \beta}$ & deformation tensor
 (Eq.~(\ref{eqn:ZA-deformation}))  \\
$\lambda_i^0$ & eigenvalue of $\partial \psi_{,\alpha}/\partial
q_{\beta}$ \\
$w_i$      & eigenvalue of $X_{\alpha \beta}$
 (Eq.~(\ref{eqn:ZA-eigenvalue})) \\ \hline
\end{tabular}
\end{table}
\vspace{-3cm}
\begin{table}
\caption{\label{tab:P-variable} Some quantities appearing in 
the pressure model.}
\begin{tabular}{c|l} \hline \hline
$\gamma$ & polytropic index in equation of state \\
$\kappa$ & proportional constant in equation of state \\
$\bm{K}$ & Lagrangian wavenumber vector \\
$K_J$    & Jeans wavenumber in Lagrangian coordinates 
 (Eq.~(\ref{eqn:Pressure-Jeans})) \\
$k_J$       & Jeans wavenumber in Eulerian coordinates \\
\hline
\end{tabular}
\end{table}

\end{document}